\documentclass[twocolumn]{aastex6}
\RequirePackage{lineno}
\usepackage{amsmath}
\usepackage{amssymb}
\usepackage{amsthm}
\usepackage{natbib,url,bm}
\usepackage{array}
\usepackage{float}
\usepackage{graphicx}
\usepackage{subfigure}
\usepackage{color}

\newcounter{ichi}
\newcounter{ni}
\newcounter{san}
\newcounter{yon}
\def\be{\begin{equation}}
\def\ee{\end{equation}}
\def\ba{\begin{eqnarray}}
\def\ea{\end{eqnarray}}

\slugcomment{draft v2}

\shorttitle{Optical emission of double pulsar formation}
\shortauthors{}

\begin{document}


\title{
Rapidly Rising Optical Transients  from the Birth of Binary Neutron Stars}


\author{Kenta Hotokezaka\altaffilmark{1}}
\author{Kazumi Kashiyama\altaffilmark{2,3}}
\author{Kohta Murase\altaffilmark{4,5,6,7}}


\altaffiltext{1}{Center for Computational Astrophysics, Flatiron Institute, 162 5th Ave, New York, NY 10010, USA}
\altaffiltext{2}{Department of Physics, the University of Tokyo, Bunkyo, Tokyo 113-0033, Japan}
\altaffiltext{3}{Research Center for the Early Universe, the University of Tokyo, Tokyo 113-0033, Japan}
\altaffiltext{4}{Department of Physics, The Pennsylvania State University, University Park, PA 16802, USA}
\altaffiltext{5}{Department of Astronomy \& Astrophysics, The Pennsylvania State University, University Park, PA 16802, USA}
\altaffiltext{6}{Center for Particle and Gravitational Astrophysics, The Pennsylvania State University, University Park, PA 16802, USA}
\altaffiltext{7}{Yukawa Institute for Theoretical Physics, Kyoto University, Kyoto 606-8502, Japan}


\begin{abstract}
We study optical counterparts of a new-born pulsar in a double neutron star system like PSR~J0737-3039A/B. 
This system is  believed to have ejected a small amount of mass of $\mathcal{O}(0.1M_{\odot})$ at the second core-collapse supernova.  
We argue that the initial spin of the new-born pulsar can be determined by the orbital period at the time when the second supernova occurs. 
The spin angular momentum of the progenitor is expected to be similar to that of the He-burning core, which is tidally synchronized with the orbital motion, and then the second remnant may be born as a millisecond pulsar.
 If the dipole magnetic field strength of the nascent pulsar is comparable to that inferred from the current spin-down rate of PSR~J0737-3039B, the initial spin-down luminosity is comparable to the luminosity of super-luminous supernovae.
We consider thermal emission arising from the supernova ejecta  driven by the relativistic wind from  such a new-born pulsar.  
The resulting optical light curves have a rise time of $\sim 10$~days and peak luminosity  of $\sim 10^{44}$~erg/s.
The optical emission may last for a month to several months, due to the reprocessing of X-rays and UV photons via photoelectric absorption.
These features are broadly consistent with those of the rapidly rising optical transients. 
The high spin-down luminosity and small ejecta mass are favorable for the progenitor of the repeating fast radio burst, FRB 121102. We discuss a possible connection between newborn double pulsars and fast radio bursts.
\end{abstract}



\section{Introduction}
The double pulsar, PSR~J0737-3039A/B, is one of the most important stellar objects for studying
compact stars and gravity~\citep{lyne2004Sci, kramer2006Sci}. 
Such a double pulsar system will eventually merge due to gravitational-wave emission,
which is one of the main targets of ground-based gravitational-wave detectors, Advanced LIGO/Virgo and KAGRA.
Binary neutron stars are also considered to produce a short gamma-ray burst (GRB) at the merger~\citep{eichler1989Nature}.  
While the orbital parameters of this system and the pulsar characteristics have been well studied, 
the formation path is still a mystery~(e.g.,~\citealt{yungelson2014LRR} for a recent review).  
The current orbital and proper motion imply that the second core-collapse supernova,
which formed the younger pulsar PSR~J0737-3039B, was an ultra-stripped supernova
associated with a very small amount of mass ejection of $\mathcal{O}(0.1M_{\odot})$~\citep{Piran(2005),dallosso2014MNRAS}.
Furthermore, \cite{vandenHeuvel2007,BP(2016)} pointed out that a majority of
the observed double neutron star systems in the Galactic plane should originate from ultra-stripped supernovae
in order to remain bound.

A small amount of ejecta suggests that supernovae associated with double pulsar formation 
is fainter than normal core-collapse supernovae if the emission powered only by 
the radioactivity of $^{56}$Ni and $^{56}$Co~\citep{tauris2013ApJ,suwa2015MNRAS,moriya2017MNRAS}. 
However, the spin-down power of new-born pulsars may be much higher than the radioactive
power. With the observed magnetic field strength of PSR~J0737-3039B and its initial spin estimated via the assumption of the tidal synchronization, the second supernova may leave a fast-rotating pulsar as a compact remnant.  
Then, the emission of supernovae associated with the double pulsar formation like PSR~J0737-3039A/B 
is likely powered by the pulsar wind, 
as considered in the pulsar-driven model for super-luminous supernovae~\citep{kasen2010ApJ,Woosley_2010}.

Recent high-cadence optical transient surveys have
discovered rapidly evolving transients (e.g.,~\citealt{drout2014ApJ,arcavi2016ApJ, tanaka2016ApJ,whitesides2017}).
Some of these transients may arise from the core collapse of an ultra-stripped star. 
For instance, \cite{arcavi2016ApJ} find rapidly rising optical transients, which have a rise time scale of
10 days or even shorter and  peak luminosities between $10^{43}$--$10^{44}$~erg/s (see also \citealt{whitesides2017}).
These luminosities are in between those of typical type II and super-luminous supernovae.
The nickel mass required to explain their peak luminosity exceeds the total ejecta mass inferred from the rise time,
suggesting that there is an extra energy source in addition to the radioactivity.

{\cite{lyubarsky2014MNRAS,CordesWasserman16,popov2016MNRAS} suggested that fast radio bursts (FRBs) can be explained as supergiant pulses arising from magnetized neutron stars. }
Such pulsar-driven supernovae became of interest in view of one the plausible candidates of FRBs. 
It has also been predicted that a bright long-term radio emission may naturally arise from synchrotron emission of 
non-thermal electrons and positrons in a pulsar wind nebula~\citep{murase2016MNRAS}.
The repeating FRB~121102 discovered by the Arecibo telescope \citep{spitler2016Nature}, whose host galaxy was recently identified following the FRB detection of {\it Very Large Array}~(VLA), is accompanied by a persistent radio counterpart that is possibly associated with the FRB source~\citep{Chatterjee2017Natur,tendulkar2017ApJ}.
A young pulsar with a high spin-down luminosity surrounded by a small amount of supernova ejecta has been argued as one of the 
plausible candidates of the radio source~\citep{kashiyama2017ApJ}.

In this work, we explore optical counterparts of newborn double pulsar systems such as PSR~J0737-3039A/B.
We estimate the ejecta mass of the second supernova and an initial spin period and magnetic field strength of the associated remnant in \S\ref{sec:parameters}. 
Then we study the optical emission leaking from the supernova ejecta powered by the pulsar wind in \S\ref{sec:opt}.
In \S\ref{sec:FRB}, we discuss the possible connection of the young binary neutron stars with
repeating FRBs.
In \S\ref{sec:con}, we conclude our results and discuss the observational implications.

%
\section{Millisecond pulsars arising from ultra-stripped progenitors in close binaries}\label{sec:parameters}
\subsection{Observational characteristics of the known double pulsar system PSR~J0737-3039A/B}
The radio observations of PSR~J0737-3039A/B provide us implications for their progenitors 
and the second core-collapse supernova, where the younger pulsar of the system was formed. 
Here we summarize some relevant parameters inferred from the observations. 
 
PSR~J0737-3039A/B is a double pulsar system of which the neutron star masses are $1.338_{-0.0007}^{+0.0007}M_{\odot}$
and $1.249_{-0.0007}^{+0.0007}M_{\odot}$ and the current orbital period is $0.102$~day
with an orbital eccentricity of $0.088$ \citep{lyne2004Sci,kramer2006Sci}. 
The current spin periods of PSR~J0737-3039A and B are $23$~ms and $2.8$~s,
respectively\footnote{PSR~J0730-3039B has not been seen from the Earth since 2008.}.
The proper motion of the system is also measured as $\sim 9$~km/s \citep{kramer2006Sci,deller2009Sci}. 
These orbital parameters and proper motion suggest that the mass ejection at the second core collapse $M_{\rm ej}$ is
only $0.1$--$0.2M_{\odot}$~(e.g., \citealt{Piran(2005),dallosso2014MNRAS}). 
This very small amount of the ejecta requires that the progenitor was an ultra-stripped star. 
\cite{BP(2016)} showed that a majority of the observed double neutron star systems in the Galactic disk 
are expected to originate from ultra-stripped progenitors (see also \citealt{vandenHeuvel2007}).

The spin-down rate of the pulsars indicates that the current magnetic
field strengths are $10^{9.8}$G and $10^{12.2}$G for
PSR~J0737-3039A and B, respectively \citep{lyne2004Sci}. 
The fast spin and weak magnetic field of PSR~J0737-3039A suggest that
this pulsar was formed first and experienced mass accretion from
the companion. Therefore, PSR~J0737-3039B is considered to be
formed at the second core collapse. 
Note that the current magnetic field strengths may be weaker than their
initial strengths because the magnetic fields might have decayed. 
We focus on  the second core-collapse event associated with the formation of PSR~J0737-3039B-like objects
throughout the paper. 

In addition to the double pulsar system PSR~J0737-3039A/B, 
PSR~J1906+0746 is likely to be the younger pulsar in a close neutron star binary with an orbital period of $0.17$~day~\citep{lorimer2006ApJ,vanleeuwen2015ApJ}. 
The magnetic field strength and the ejecta mass at the second core collapse are estimated as
$10^{12.2}$~G and $\sim 0.8M_{\odot}$, respectively \citep{lorimer2006ApJ,BP(2016)}. 
Note, however, that this ejecta mass is estimated based on the upper limit on the proper motion $\lesssim 400$~km/s
of this system, so that it could be smaller~\citep{vanleeuwen2015ApJ}.

\subsection{A progenitor scenario}
We consider the following scenario of the double neutron star formation (see 
\citealt{tauris2013ApJ,tauris2015MNRAS,suwa2015MNRAS} for the evolution of ultra-stripped progenitors):
\begin{enumerate}
\item A He star with a mass $\sim 3M_{\odot}$ orbits with an orbital period of $\sim 0.1$ day
around a neutron star that is formed at the first supernova.
In such a system, the He star is quickly tidally synchronized with the orbital motion, as will be discussed later.
\item After the core He-burning finishes, the tidal torque on the star is significantly reduced since the convective core contracts.
While the angular momentum of the envelope is lost due to the wind mass loss during this stage, 
the angular momentum of the core may not be significantly transferred to the envelope (\citealt{hirschi2005A&A,yoon2005A&A,meynet2007A&A}).
\item At the end of the stellar evolution, the second supernova occurs, where a small amount of mass
$\mathcal{O}(0.1M_{\odot})$ is ejected. The newly formed pulsar rotates with the angular momentum of the core prior to the collapse.
This rotational energy is the source of a supernova powered by a pulsar wind discussed 
in the following sections.
\end{enumerate}

Here we make two simple but reasonable assumptions. 
First, the spin angular momentum and mass of the core just prior to the core collapse is assumed to be equal to those of the He-burning core. 
Second, we assume a circular orbit until the core collapse.
Then, the initial spin period of PSR~J0737-3039B, $P_{s,{\rm ns}}$, can be estimated through the angular momentum conservation:
\begin{eqnarray}
P_{s,{\rm ns}}  & = & \left(\frac{r_{g,\,{\rm ns}}R_{\rm ns}}{r_{g,\,c}R_c}\right)^2 P_{\rm orb}, \notag \\
& \approx &  {0.8~{\rm ms}~\left(\frac{r_{g,\,c}^2}{0.075}\right)^{-1}
\left(\frac{r_{g,\,{\rm ns}}^2}{0.4}\right)
\left(\frac{R_{c}}{0.13R_{\odot}} \right)^{-2} }  \notag \\ 
& & {\times \left(\frac{R_{\rm ns}}{11{\rm km}} \right)^{2}
\left(\frac{P_{\rm orb}}{0.12~{\rm day}}\right), }
\end{eqnarray}
where $R_{c}$ and $R_{\rm ns}$ are the radius of the core of the He star and
of the new-born pulsar,
$r_{g,c}$ and $r_{g,\,{\rm ns}}$ are their non-dimensional
gyroradii \citep{kushnir2016MNRAS,lattimer2001ApJ}, 
$P_{\rm orb}$ is the orbital period at the second core collapse.
Here we have used the semi-major axis at the second collapse of
$\sim 10^{11}$~cm~\citep{beniamini2016ApJ}, inferred from the fact that the second core-collapse
supernova occurred at $\sim 50$~Myr ago, corresponding to  
the spin-down age of PSR~J0737-3039B.
Note that the above estimate of the pulsar's spin frequency is 
the maximal one since we have assumed that the core does not lose 
its angular momentum in the post He-burning phase.  
For instance, the initial spin period of the pulsar becomes $\sim 8$~ms if 
$90\%$ of the core's spin angular momentum is lost.
The initial spin energy of the new-born pulsar is
\begin{eqnarray}
E_{s} & = & \frac{1}{2}r_{g,\,{\rm ns}}^2 R_{\rm ns}^2 M_{\rm ns}\left( \frac{2\pi}{P_{s,{\rm ns}}} \right)^2, \notag \\
 & \approx &  {3.5\cdot 10^{52}\,{\rm erg}\,
\left(\frac{r_{g,\,{c}}^2}{0.075}\right)^{2}
\left(\frac{r^2_{g,\,{\rm ns}}}{0.4}\right)^{-1} 
\left(\frac{M_{\rm ns}}{1.3M_{\odot}} \right)} \notag \\
& & {\times \left(\frac{R_{\rm ns}}{11\,{\rm km}} \right)^{-2}
\left(\frac{R_{c}}{0.13R_{\odot}} \right)^{4}
\left(\frac{P_{\rm orb}}{0.12\,{\rm day}} \right)^{-2},}
\end{eqnarray}
where $M_{\rm ns}$ is the mass of the new-born pulsar.
Note that this energy is much larger than 
the explosion energy of  a typical supernova of $\sim 10^{51}$~erg.

The assumption of the synchronization during the core He-burning phase is justified as follows.
The tidal synchronization of the progenitor star occurs on the synchronization time \citep{zahn1975A&A,zahn1977A&A,kushnir2017MNRAS}:
\begin{eqnarray}
t_{\rm sync} \approx 300~{\rm yr}~q^{-5/6}\left(\frac{1+q}{2} \right)^{2}
\left(\frac{r_g^2}{0.075}\right)
\left(\frac{P_{\rm orb}}{0.12~{\rm day}} \right)^{17/3} \notag \\ 
\times \left(\frac{R_*}{0.43R_{\odot}} \right)^2
\left(\frac{R_c}{0.13R_{\odot}} \right)^{-9}
\left(\frac{M_*}{3M_{\odot}} \right)
\left(\frac{M_c}{1.4M_{\odot}} \right)^{4/3},\label{sy}
\end{eqnarray}
where $q$ is the mass ratio of the progenitor star to the companion neutron star, 
$R_*$ and $M_*$
are the radius and mass of the He star, and $M_c$ is the core mass. This 
timescale is much shorter than  the lifetime
and the timescale of the wind angular-momentum loss of a He star. Thus, the progenitor star is tidally synchronized
during the He-star phase. {It can be also shown that the circularization of the orbit occurs during the He-burning phase 
due to the tidal torque \citep{zahn1977A&A}.}
In the post He-burning phase, the core radius shrinks, and hence the tidal synchronization is much less 
efficient so that the progenitor star likely leaves the synchronization state (see Eq. \ref{sy} for the strong dependence of the synchronization time on the convective core radius). 
The situation is somewhat complicated for binaries with
longer orbital periods. For instance, double neutron star progenitors with a gravitational-merger time of
$t_{\rm GW}=10$~Gyr, corresponding to $P_{\rm orb}=0.65$~day, have $t_{\rm sync}\approx 3$~Myr,
which is of order of the timescale of the stellar evolution.

Note that the synchronization time scale is quite sensitive to the structure of the progenitors so that
the stellar evolution calculation is needed for 
detailed studies
(see, e.g.,~\citealt{tauris2015MNRAS} for a stellar evolution study of double neutron stars).
Furthermore, the spin angular momentum of the core may be lost during the late stage of the stellar evolution. 
The final spin angular momentum of the core depends on the efficiency of angular momentum losses 
from the star due to the wind and that from the stellar core to the envelope. The former depends on the 
anisotropy and the magnetization of the wind (see, e.g., \citealt{meynet2007A&A,ud2009MNRAS}).
The latter occurs through the angular momentum transport due to the magnetic field and meridional circulation. Note that
\cite{hirschi2005A&A,yoon2005A&A,meynet2007A&A} showed that the angular momentum exchange 
between the core and envelope may not be efficient in the post core He-burning phase.
We do not go into these issues in this work and will address them in a future work.

\section{Optical emission from pulsar-driven ultra-stripped supernovae}\label{sec:opt}
A new-born pulsar of a binary neutron star system launches
a relativistic wind, which interacts with the
supernova ejecta  surrounding the wind.
The forward shock in the supernova ejecta and
the reverse shock in the wind~(the wind termination shock) are formed. 
The internal energy of the supernova ejecta is increased
via the shock dissipation and  the reprocess of non-thermal radiation from
the pulsar wind. 
Here we consider an optical transient arising from such a system (see, e.g., \citealt{kotera2013MNRAS,yu2014,kasen2016ApJ,kashiyama2016ApJ}).

\subsection{Dynamical and thermal evolution of the system}
Pulsar winds are powered by the pulsar's spin-down energy with
a luminosity \citep{gruzinov2005PhRvL,spitkovsky2006ApJ}: 
\begin{eqnarray}
L_{\rm sd}  & \approx & \frac{\mu^2}{c^3}\left(\frac{2\pi}{P_{s,{\rm ns}}}\right)^4,\notag \\ 
& \approx & 1.4\cdot 10^{44}~{\rm erg/s}~\left( \frac{B}{10^{12.2}{\rm G}}\right)^2 \label{Lsd}
\left(\frac{P_{\rm s,ns}}{0.8~{\rm ms}} \right)^{-4},
\end{eqnarray}
where $\mu=BR_{\rm ns}^3/2$, $B$ is the dipole magnetic field strength at the neutron star surface
at the dipole pole,
and we have assumed that the dipole axis is aligned to the rotational axis of the star.
{Note that the formula shown by \cite{gruzinov2005PhRvL,spitkovsky2006ApJ} gives a spin-down luminosity 
even larger than Eq. (4) by a factor of $\leq2$ when the dipole axis is misaligned with the rotation axis}.
Here we have also assumed that the current strength of the magnetic field  
of PSR~J0737-3039B is a typical value of new-born pulsars in close double neutron star systems. 
The spin-down luminosity is constant until the spin-down time: 
\begin{eqnarray}
t_{\rm sd}  & = & \frac{E_s}{L_{\rm sd}}, \notag \\
& \approx & 8~{\rm yr}~\left( \frac{B}{10^{12.2}{\rm G}}\right)^{-2} \label{tsd}
\left(\frac{P_{\rm s,ns}}{0.8~{\rm ms}} \right)^2,\\ \nonumber
\end{eqnarray}
After $t_{\rm sd}$, the spin-down luminosity declines as $\propto t^{-2}$.
Note that we focus on the evolution of the system and its optical emission
up to $100$ days after the explosion, and hence, the spin-down luminosity 
is constant for our fiducial parameters.

The energy flux of the pulsar wind is carried by relativistic particles and magnetic fields:
\begin{eqnarray}
L_{\rm sd}  & =  &  L_{e^{+}e^{-}} + L_{B}, \\
&\equiv & (1+\sigma)L_{e^{+}e^{-}}, \nonumber
\end{eqnarray}
where $L_{e^{+}e^{-}}$ and $L_B$ are the luminosities of electrons/positrons and magnetic fields, respectively. 
The ratio of the luminosities between these two components are traditionally denoted by a parameter $\sigma$.  
The out-going energy flux near the pulsar is dominated by the magnetic filed, i.e., $\sigma \gg1$. 
Then the magnetic energy is assumed to be converted to the particles' energy due to either magnetic reconnection 
or plasma/magnetrohydrodynamical instabilities around the wind termination shock. 
As in the previous works for pulsar-driven supernovae~\citep[e.g.,][]{murase2015ApJ,kashiyama2016ApJ} we consider the limit $\sigma\ll 1$, i.e, $L_B\ll L_{e^{+}e^{-}}$, in the nebula just outside the wind termination shock.  
This is motivated by the one-zone modeling of the Crab pulsar wind nebula (see, e.g.,~\citealt{rees1974MNRAS,kennel1984ApJ}). 

Thermal emission emerging from the supernova ejecta powered by a pulsar wind is described by the first law of thermodynamics of
the ejecta \citep{arnett1979ApJ,kasen2010ApJ}:
\begin{eqnarray}
\frac{dE_{\rm int}}{dt} = -\frac{E_{\rm int}}{t} - L_{\rm rad} + \dot{Q},\label{therm}
\end{eqnarray}
where $E_{\rm int}$ is the internal energy, 
$L_{\rm rad}$ is the thermal radiation cooling rate, and $\dot{Q}$ is the heating rate of the ejecta.
Here we describe the thermal radiation cooling as $L_{\rm rad}\approx E_{\rm int}/t_{\rm rad}$,
where $t_{\rm rad}=3\xi \kappa M_{\rm ej}/4\pi v_{\rm ej}ct$ is the 
photon diffusion time scale at a given time, $\kappa$ is the opacity of the ejecta
 to the thermal photons, $v_{\rm ej}$ is the typical  ejecta velocity,
 $\xi$ is a geometrical factor depending on the ejecta profile.

We turn now to discuss the heating rate $\dot{Q}$ and 
the dynamics of the ejecta. As long as the supernova ejecta is sufficiently optically thick, 
the spin-down power injected to the ejecta forms a forward shock in the ejecta irrespective of the value of $\sigma$.
During this phase, roughly a half of the spin-down energy is converted to the internal energy and another half is
converted to the kinetic energy. 
The energy dissipation at the forward shock is efficient until the time when the thermal radiation efficiently escapes from the system\footnote{The synchrotron cooling time of relativistic electrons and positrons injected in the wind nebula is faster than the dynamical time 
(see, e.g.,~\citealt{metzger2014MNRAS,murase2015ApJ}). Therefore the pressure in the wind nebula is dominated by radiation.
Once the radiation starts to diffuse out from the system, the pulsar wind may not push the supernova ejecta efficiently.}
or the total injected energy becomes comparable to the ejecta's initial kinetic energy.
The former occurs when the diffusion time of thermal photons in the ejecta is comparable to the expansion time: 
\begin{eqnarray}
t_{\rm diff}  & =& \left( \frac{3\xi \kappa M_{\rm ej} }{4\pi c v_{\rm ej} }\right)^{1/2}, \\
& \approx & 5\,{\rm day}\,\xi^{1/2} \left(\frac{\kappa}{0.1\,{\rm cm^2/g}} \right)^{1/2}
{\small
\left(\frac{M_{\rm ej}}{0.1~M_{\odot}} \right)^{3/4}
\left(\frac{E_{\rm sn}}{10^{50}\,{\rm erg}} \right)^{-1/4},\notag}
\end{eqnarray}
where we use the opacity as the sum of the electron scattering of partially ionized plasma and 
the bound-bound absorption, and $\xi$ is chosen to be unity. 
Here the supernova kinetic energy $E_{\rm sn}=10^{50}$~erg corresponds to
the initial ejecta velocity of $v_{{\rm ej},0}\approx 0.03c$ and
the ejecta mass of $M_{\rm ej}\approx 0.1M_{\odot}$.
The latter occurs at the sweep-up time $t_{\rm sw}$, which is estimated by
\begin{eqnarray}
\eta M_{\rm ej} v_{\rm ej}^2 & \approx &  \int^{t_{\rm sw}} L_{\rm sd}dt 
\approx  t_{\rm sw} L_{\rm sd},
\end{eqnarray}
where $\eta$ is an order-of-unity parameter that depends on the ejecta's structure (\citealt{suzuki2017MNRAS}).
The sweep-up time is thus
\begin{eqnarray}
t_{\rm sw} & \approx& 8~{\rm day}~\eta \left( \frac{B}{10^{12.2}{\rm G}}\right)^{-2} \notag \\ 
& & \times \left(\frac{P_{\rm s,ns}}{0.8~{\rm ms}} \right)^{4} \left(\frac{E_{\rm sn}}{10^{50}\,{\rm erg}} \right). 
\end{eqnarray}
The supernova ejecta expands with a constant velocity until $t\sim t_{\rm sw}$.
After $t_{\rm sw}$, the injected energy from the pulsar wind exceeds $E_{\rm sn}$ and
the ejecta velocity increases with time if $t_{\rm diff}>t_{\rm sw}$.
%
We describe the shock heating rate as
\begin{eqnarray}
\dot{Q}_{\rm sh}(t) 
& \approx & \frac{1}{2}L_{\rm sd}~~~~~~~~\,({\rm for}~~ t \leq t_{\rm sw},~t_{\rm diff} ). 
\end{eqnarray}
For $t > t_{\rm sw}$,
the shock proceeds in the outer part of the ejecta, where the  density gradient
is rather steep \citep{chevalier1989ApJ}. In such a region, the shock heating is less efficient compared
to those due to the irradiation by non-thermal photons produced in the pulsar wind nebula as discussed later.
 Here  we neglect the shock heating for $t>t_{\rm sw}$ (see \citealt{kasen2016ApJ}
 for the shock heating rate in the ejecta).
Furthermore, $t_{\rm sw}$ is longer than
$t_{\rm diff}$ for the parameters considered in this paper (see Table \ref{tab:para}).
In such cases, the forward shock breaks out by $t_{\rm sw}$.
{After the shock breaks out, the energy injection into the ejecta is weaken and
radiative cooling becomes quite efficient. As a result, the ejecta is slowly accelerated due to
the adiabatic expansion and the momentum injection. These may result in the increase 
of the ejecta velocity by $\lesssim 50\%$ at $100$~days. However, we simply assume that the ejecta velocity 
is constant with time in this work because we do not solve the X-ray and UV absorptions of the ejecta 
at late times properly (see the following discussion).  }


\begin{table*}[t]
\begin{center}
\caption[]{Parameters of theoretical light curves}
\label{tab:para}
\begin{tabular}{lccccccc}
\hline \hline
Event & redshift  & $\kappa\,[{\rm cm^2/g}]$ & $M_{\rm ej}\,[M_{\odot}]$  & $v_{\rm ej}\,[c]$ & $E_{\rm sn}\,$[erg] &$L_{\rm sd}\,[{\rm erg/s}]$ & $f_{\rm_{X-UV},bf}$  \\ \hline
PTF10iam & $0.109$ & $0.1$ & $0.1$ & $0.07$ & $4.4\cdot 10^{50}$ & $2.4\cdot 10^{44}$& $0.9$\\
SNLS04D4ec & $0.593$ & $0.1$ & $0.1$ & $0.09$ & $7.3\cdot 10^{50}$ & $2.9\cdot 10^{44}$ & $0.05$\\
SNLS05D2hk & $0.699$ & $0.1$ & $0.1$ & $0.06$ & $3.2\cdot10^{50}$ & $3.2\cdot10^{44}$ & $0.1$\\
SNLS06D1hc & $0.555$ & $0.1$ & $0.1$ & $0.07$ & $4.4\cdot10^{50}$ & $2.6\cdot10^{44}$ & $0.1$\\
\hline \hline 
\end{tabular}
{ \\
The redshifts are taken from \cite{arcavi2016ApJ}.
}
\end{center}
\end{table*}

The heating due to the reprocessing of non-thermal photons produced in the nebula can be efficient even at late times.
Here we treat these processes in a simple way as follows. 
At early times, electromagnetic cascades proceed in the saturation regime, leading to a flat energy spectrum up to $\sim1$~MeV~\citep{metzger2014MNRAS}. At later times, the spectrum depends on the seed photon spectra, but it can roughly be estimated to be a flat spectrum from $\sim 1$~eV to $\sim 0.1$~TeV while the supernova emission continues, which is expected based on more detailed calculations (e.g.,~\citealt{murase2015ApJ,murase2017}). 
High-energy $\gamma$ rays ($\gtrsim 1$~MeV)~heat up the ejecta through the Compton scattering and the pair production process. 
X-ray and UV photons are absorbed and heat up the ejecta through the photoelectric~(bound-free) absorption unless the ejecta are fully ionized. 
Here we describe the heating rate as
\begin{eqnarray}
\dot{Q}_{\rm rad}(t) \approx \left(f_{\gamma} + f_{\rm_{X-UV},bf}\right) L_{\rm sd},
\end{eqnarray}
where $f_{\gamma}$ and $f_{\rm_{X-UV},bf}$ are the heating
efficiencies of $\gamma$ rays and X-ray and UV photons  to the spin-down luminosity,
respectively. We calculate the frequency dependent heating efficiency 
of $\gamma$ rays at each time:
\begin{eqnarray}
f_{\gamma}  (t) = \frac{\int_{\nu_{\rm min}}^{\nu_{\rm max}} \frac{d\nu}{\nu} \,{\rm min}(K_{\gamma,\nu}{\tau}_{\gamma,\nu},1)}{\int_{1\,{\rm eV}}^{1\,{\rm TeV}}\frac{d\nu}{\nu}},
\end{eqnarray}
where  the 
frequency range of $\gamma$ rays is $(h\nu_{\min},h\nu_{\rm max})=(10\,{\rm keV},1\,{\rm TeV})$, and
$h$ is the Planck constant. 
Here, ${\tau}_{\gamma,\nu}$ is the optical depth of the ejecta to $\gamma$ rays and $K_{\gamma,\nu}$ is the photon inelasticity at a given frequency, where the Klein-Nishina cross section and the cross section for the Bethe-Heitler pair production in the field 
of a carbon nucleus are taken into account~\citep{murase2015ApJ,chodorowski1992ApJ}. 
Note that the coefficient of 
the $\gamma$-ray optical depth depends on the density profile of the ejecta. Here
we simply assume a density profile to be constant with the radius.
Adopting a realistic density profile may result in different ejecta mass and velocity estimates by a factor of a few. 

The heating efficiency of X-ray and UV photons ($\sim 10\,$eV to $\sim 10\,$keV) are 
somewhat more difficult to estimate because it depends  on the ionization state of the ejecta. 
The value of $f_{\rm_{X-UV},bf}$ is limited by the energy fraction of 
photons with energies from $\sim 10\,$eV to $\sim 10\,$keV, and hence,  $f_{\rm_{X-UV},bf}\lesssim 1/4$.  
In this work, we  assume $f_{\rm_{X-UV},bf}$
to be constant  with time and we determine  $f_{\rm_{X-UV},bf}$ such that
the late-time tail of the theoretical light curve reproduces the observed light curves (see Table \ref{tab:para}).
Note that the photoelectric absorption is efficient until the supernova ejecta are fully ionized.
The ionization break occurs at different times for different frequencies (see \citealt{metzger2014MNRAS}).
For instance, the ionization break-out time for iron  can be estimated as
\begin{eqnarray}
t_{\rm bo}& \approx & 100~{\rm day}~\left(\frac{M_{\rm ej}}{0.1M_{\odot}} \right) 
\left(\frac{v_{\rm ej}}{10^9\,{\rm cm/s}} \right)^{-3/2}
\left(\frac{T}{10^5\,{\rm K}} \right)^{-0.4} \\ \nonumber
& & \times \left(\frac{X_{\rm Fe}}{0.35} \right)^{1/2}
 \left(\frac{L_{\rm sd}t}{1.7\cdot 10^{51}\,{\rm erg}} \right)^{-1/2}
\left(\frac{Z}{26}\right)^{3/2},
\end{eqnarray}
where $X_{\rm Fe}$ is the mass fraction of iron of the ejecta \citep{suwa2015MNRAS}.
The ionization break time of the lighter elements is shorter,
 thereby  $f_{\rm_{X-UV},bf}$ may decrease with time on a timescale of $\sim 100$~days
 in a realistic situation. {Note that Rayleigh-Taylor instability may grow and affect significantly the
 ejecta structure at late times ($t\gtrsim t_{\rm sw}$, see, e.g., \citealt{suzuki2017MNRAS}). 
 This affects the absorption probability of X-ray and UV photons
 and the estimate of the ionization break-out time.  }

In summary, a supernova ejecta powered by a pulsar wind is heated by 
(i) the forward shock in the ejecta until $t\approx {\rm min}(t_{\rm diff},t_{\rm sw})$,
(ii) the Compton scattering and pair production process to $\gamma$ rays emitted by non-thermal electrons and positrons in 
the pulsar wind nebula, which last until the ejecta becomes optically thin to $\gamma$ rays,
and (iii) the photoelectric absorption of X-ray and UV photons, which likely lasts until $100$~days.
This injected energy into the ejecta is cooled by the adiabatic expansion and radiative losses.
Note that we neglect the radioactive heating as it is expected to be much smaller than the spin-down luminosities 
for our fiducial parameters.
The resulting bolometric light curve of the thermal radiation is obtained as
\begin{eqnarray}
L_{\rm rad}(t) \approx
\frac{\exp\left(-\frac{t^2}{2t_{\rm diff}^2}\right)}{t_{\rm diff}^2}\int^{t}dtt \dot{Q}(t)\exp\left(\frac{t^2}{2t_{\rm diff}^2}\right),
\end{eqnarray}
where we have assumed that the ejecta are not accelerated significantly, 
and the initial internal energy of the supernova ejecta does not contribute to the radiation.
The latter is justified  because
the adiabatic cooling is  efficient for compact progenitors.

\begin{figure}
\includegraphics[scale=0.7, bb=50 50 400 280 ]{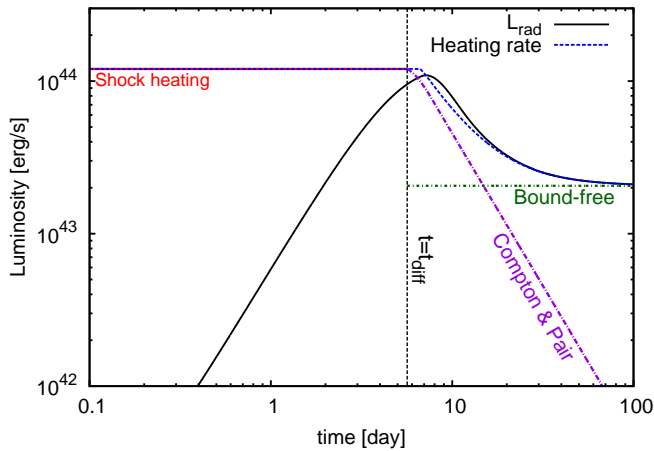}
\caption{
The bolometric light curve~(black solid curve) and heating rates. The ejecta mass,
initial kinetic energy of the ejecta, and
the spin-down luminosity of the new-born pulsar are 
chosen to be $0.1M_{\odot}$, $4.4\cdot 10^{50}\,$erg, and $2.4\cdot10^{44}$~erg/s, respectively.
The heating rate due to the shock, the $\gamma$-ray heating through the Compton scattering
and pair production, and bound-free absorption are
shown as a red solid, purple short dot-dashed, and green long dot-dashed curve, respectively.
Also shown as a vertical  line is the diffusion time of thermal photons.
}
\label{fig:Lbol}
\end{figure}

Figure \ref{fig:Lbol} shows the bolometric light curve of the thermal emission $L_{\rm rad}$
and the total heating rate. The heating rate of each process is also depicted. 
Here we use the parameters of PTF10iam shown in Table \ref{tab:para}.
The bolometric light curve arises on the diffusion time scale and has a peak luminosity of $\approx L_{\rm sd}/2$.
It declines fast after $t_{\rm diff}$ to $20$~days, where the Compton scattering and pair production dominate the 
heating rate and $\gamma$ rays start to leak from the ejecta on this time scale.
After $20$~days, the bound-free absorption to X-ray and UV photons dominates the heating rate,
which contributes to the long-lasting tail of the bolometric light curve. 

\subsection{Optical light curves}

We calculate bolometric light curves of the supernova thermal radiation 
using Eq.~(\ref{therm}). In order to obtain the light curve 
at a given frequency, we assume the black-body spectrum
with a temperature given by $T_{\rm eff}=(L_{\rm rad}/4\pi \sigma_{\rm SB}r^2)^{1/4}$,
where $\sigma_{\rm SB}$ is the Stefan-Boltzmann constant \citep{arnett1979ApJ}.
For instance, the effective temperature at the peak time $\approx t_{\rm diff}$ 
with the peak luminosity $\approx L_{\rm sd}/2$ is
\begin{eqnarray}
T_{\rm eff} \approx 4\cdot 10^4~{\rm K}~\left( \frac{B}{10^{12.5}{\rm G}}\right)^{0.5}
\left(\frac{P_{\rm s,ns}}{1~{\rm ms}} \right)^{-1} \notag \\
\times \left(\frac{v_{\rm ej}}{0.03~c} \right)^{-1/2}
\left(\frac{t_{\rm diff}}{5~{\rm day}} \right)^{-1/2}.\\ \nonumber
\end{eqnarray}
We expect  bright thermal radiation in the UV to optical bands after the peak time.
The black-body assumption cannot be justified once the photosphere shrinks significantly.
This occurs shortly after the peak of the bolometric light curve. 
Then the emission is dominated by the nebula emission rather than the photospheric emission.
However, the calculation of the nebular emission requires detailed treatments on the radiative transfer, 
which are beyond the scope of this paper.

\begin{figure*}[t]
\includegraphics[scale=0.7,bb=50 50 400 280]{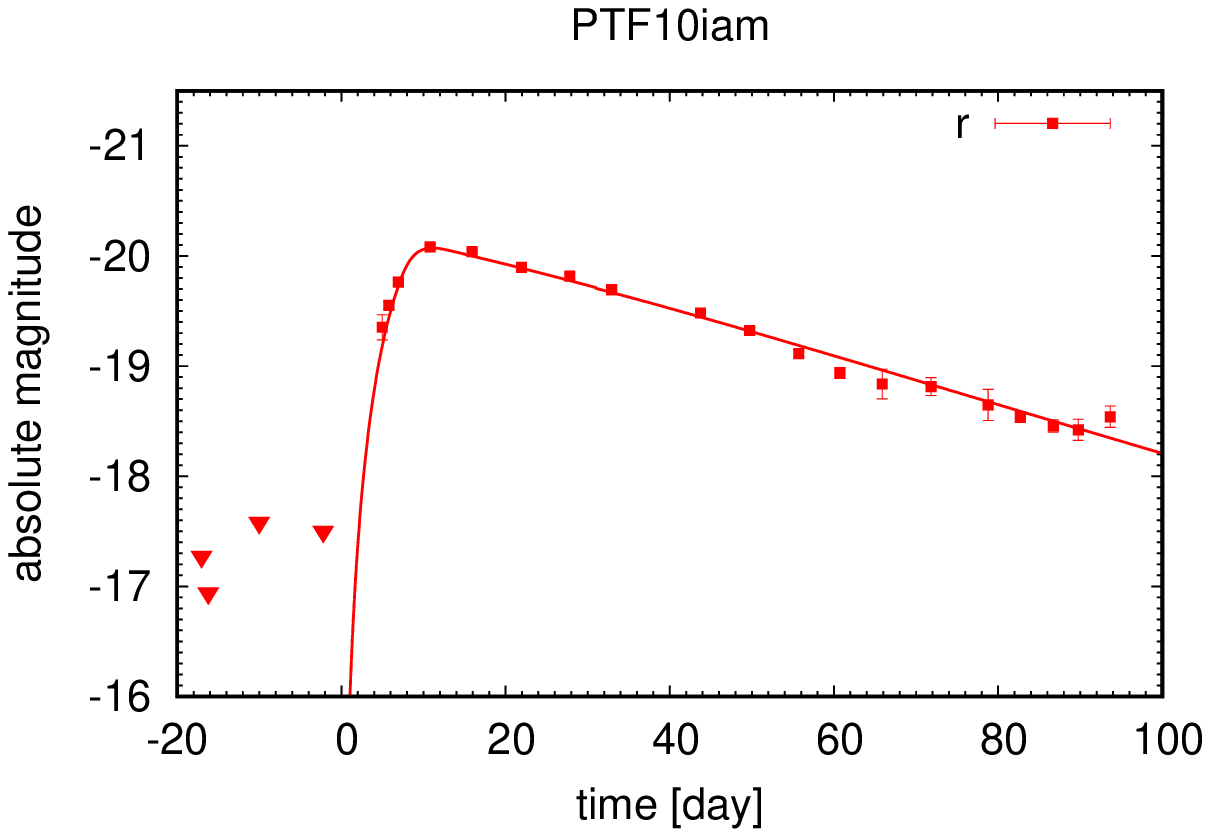}
\includegraphics[scale=0.7,bb=50 50 400 280]{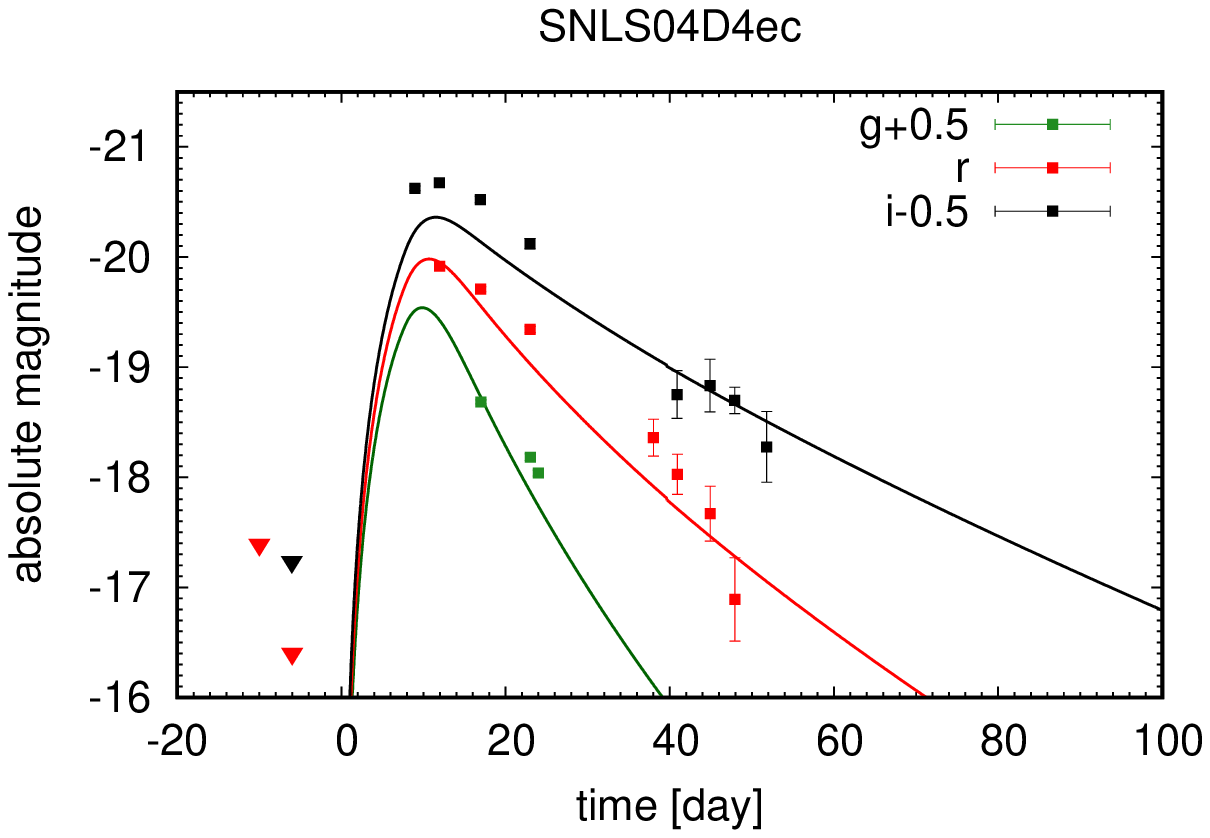}\\
\includegraphics[scale=0.7,bb=50 50 400 280]{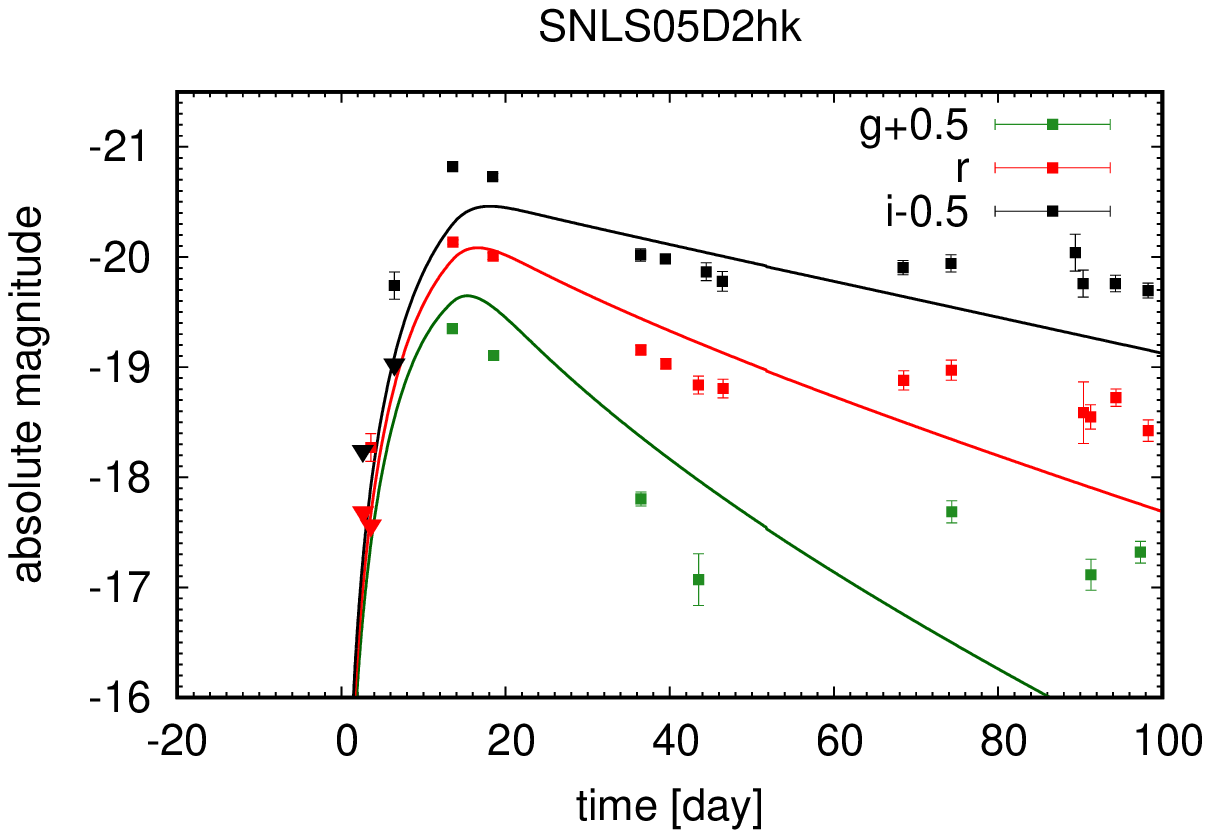}
\includegraphics[scale=0.7,bb=50 50 400 280]{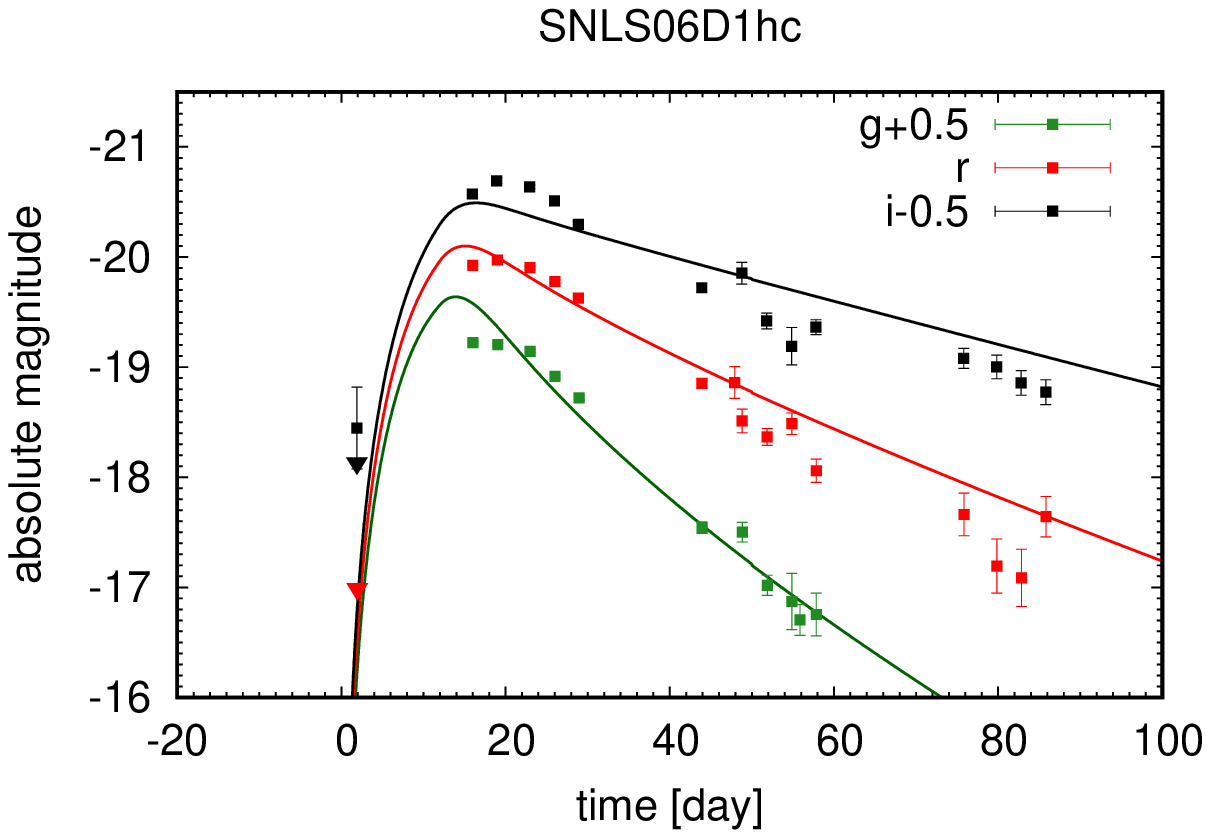}\\
\caption{
Absolute magnitude of the optical emission from a supernova ejecta interacting with 
a new-born pulsar wind and the observed  data of the rapidly rising optical transients 
taken from \cite{arcavi2016ApJ}. The detections and the 3-sigma upper limits 
are depicted as squares and triangles, respectively.
Here we take the effect of the cosmological redshift on the observed time and 
observed flux into account for
the theoretical curves. 
The parameters of the theoretical curves used for each event are listed in Table \ref{tab:para}.
}
\label{fig:L}
\end{figure*}

Figure \ref{fig:L} shows the light curves of the thermal emission
arising from the pulsar-driven supernova with small ejecta mass.
Also shown are the observed light curves of rapidly rising optical transients \citep{arcavi2016ApJ}. 
The peak luminosity and rise time scale are basically determined by $L_{\rm sd}$ and $\kappa M_{\rm ej}/v_{\rm ej}$, respectively.
The slope of the tail depends on $f_{\rm_{X-UV},bf}$ and $v_{\rm ej}$.
The parameters used for each event are listed in Table \ref{tab:para}. 
These parameter ranges are inferred from the formation scenario of double pulsar systems 
like PSR~J0737-3039A/B, as described in the previous section.  
Note that we have four independent parameters, $\kappa M_{\rm ej}$, 
$v_{\rm ej}$, $L_{\rm sd}$, and $f_{\rm_{X-UV},bf}$, to generate the theoretical light curves.
It is worthy noting that the observed data are reproduced with the ejecta's kinetic energies
of $3$--$8\cdot 10^{50}$~erg, which are consistent with 
the results of a hydrodynamical simulation of ultra-stripped supernovae \citep{suwa2015MNRAS}.
Around $100$ days, the ejecta temperature becomes somewhat low $\sim 3000$~K,
where atoms with low ionization energy, e.g, iron, are not fully recombined \citep{kleiser2014MNRAS}.
The heating efficiency of the photoelectric absorption
$f_{\rm_{X-UV},bf}$ is $0.05$ to $0.1$,  corresponding to that roughly less than
a half of the energy in X-ray and UV photons are thermalized.
Note that, however, the back-body assumption may  not be a good approximation
at late times, so that the values of $f_{\rm_{X-UV},bf}$ derived via the light curve fitting 
is physically less meaningful. 


In summary, an optical counterpart of the double pulsar formation like PSR~J0737-3039A/B, i.e., 
an ejecta mass of $\sim 0.1M_{\odot}$ and a pulsar's initial spin-down luminosity of $\sim 2\cdot 10^{44}$~erg/s,
has a fast rise time, bright peak luminosity, and long-lasting tail, which broadly
agree with the observed light curves of the rapidly rising optical transients \citep{arcavi2016ApJ}.
The long-lasting energy injection to the ejecta due to the photoelectric absorption plays a crucial role to produce the
long-lasting tail of the light curves. 
In order to reproduce the observed light curves,
one needs a spin-down luminosity of $\gtrsim 2\cdot 10^{44}$~erg/s and a spin
down time of $\gtrsim 0.3$~yr, respectively. 
These conditions give upper limits on the initial spin period and
magnetic field strength as $\sim 3$~ms  and $10^{13.5}$~G, respectively.

\subsection{Rates and diversity}
The rate of ultra-stripped supernovae powered by a new-born pulsar in a double pulsar system 
is expected to be $\sim 0.1\%$ of that of normal core-collapse supernovae. 
This number is estimated from the population of double neutron stars in the Galaxy~\citep{kalogera2004ApJ,kim2015MNRAS},
as well as the rate of short GRBs (with a correction of the beaming factor;~\citealt{wanderman2015MNRAS}). 
While the rate of rapidly rising optical transients is still unknown, the inferred rate of $\sim{10}^{2}~{\rm Gpc}^{-3}~{\rm yr}^{-1}$ looks consistent with this rate. 
{In this work, we focus on the transients reported by \cite{arcavi2016ApJ} and we propose that they are ultra-stripped supernovae
occurring in very close binaries.
The rapidly-evolving transients reported by  \cite{drout2014ApJ}, which are  fainter than those in \cite{arcavi2016ApJ}, may also 
arise from the birth of a double pulsar system with different parameters, e.g., larger orbital separations. 
Their event rate  is estimated as $\sim 5\cdot 10^3~{\rm Gpc}^{-3}~{\rm yr}^{-1}$. Given the large uncertainties in the estimates of the
birth rate of double pulsar systems, these transients might be explained as supernovae associated double pulsars.} 

The formation of double pulsars likely has variations in
the orbital period at the second core collapse and the strength of
the magnetic field. Note also that, as we mentioned earlier, 
the initial spin period of pulsars depends on
the mass loss history of the post He-burning phase. 
Therefore we expect there to be
a broad range of the peak luminosities and rise times (see \citealt{kashiyama2016ApJ} for 
a study with a wide range of parameters). 
For instance,
the spin-down luminosity of a pulsar in a binary with
an orbital period of $0.65$~days, corresponding to a merger time of $10$~Gyr,
is $\sim 3\cdot 10^{41}$~erg/s, if the progenitor star is tidally synchronized during the core He-burning phase.
In such a case, the radioactivity of $^{56}$Ni may provide more energy 
than the pulsar wind at the peak time of the light curve, and hence, the 
peak luminosity is much fainter. Such population may explain some of the observed faint ultra-stripped supernovae
and calcium-enriched gap transients~\citep{moriya2017MNRAS}.

\section{Connection between fast radio bursts and double pulsar systems?}\label{sec:FRB}

We now turn to discuss a scenario that the new-born pulsars in close double neutron stars are the progenitors of FRBs. 
Young neutron star systems have been intensively investigated as the FRB sources~\citep[e.g.,][]{Popov&Postnov10,lyubarsky2014MNRAS,CordesWasserman16,popov2016MNRAS,Connor+16,murase2016MNRAS}, 
and even more so for the repeating FRB 121102 after the discovery of its host galaxy and persistent radio counterpart.
Several authors claim that a high spin-down luminosity and/or small ejecta mass are favored to explain the observed characteristics of FRB 121102~\citep{kashiyama2017ApJ,metzger2017,kisaka2017,katz2017,dai2017ApJ,Piro_Burke-Spolaor_17,waxman2017}. 
In particular, \cite{kashiyama2017ApJ} showed that the persistent radio counterpart is consistent with the radio emission arising from a pulsar wind nebula in an ultra-stripped supernova remnant with $M_{\rm ej} \sim 0.1 \ M_\odot$, powered by a new-born pulsar 
with $B \sim 10^{12}$--$10^{13} \ \rm G$, $P_{\rm s, ns} \lesssim$ a few ms, and an age of $\sim 10 \ \rm yr$.
The above parameters are in accord with those of new-born pulsars in close double neutron stars.
 
 {The new-born pulsar scenario of the repeating FRB suggests that the pulsar-wind nebula emits radiation in
 a broad range of wavelengths, e.g., optical, X ray, $\gamma$ ray. The spatially resolved optical-IR emission 
 around  FRB 121102  is consistent with the emission of an active star forming region \citep{tendulkar2017ApJ,bass2017ApJ}. 
 The discovery of such radiation in X ray and $\gamma$ ray will be strong evidence for supporting this scenario.
 However, X-ray emission is not  detected by XMM-Newton and Chandra so far and the derived upper limit is $\nu F_{\nu}\sim 3\cdot 10^{-15}\,{\rm erg\,cm^{-2}\,s^{-1}}$ \citep{Chatterjee2017Natur}. This upper limit is still consistent
 with the model that the persistent emission of the repeating FRB has a spectral shape similar to the Crab's one \citep{Chatterjee2017Natur} and
 that of theoretical modeling of a young pulsar wind nebular emission \citep{murase2016MNRAS,murase2017}.
 }

The population of repeating FRB sources can be estimated as follows.
The rate of repeating FRBs inferred from the survey is $5.1^{+17.8}_{-4.8}\cdot 10^{4}~{\rm sky^{-1}\,day^{-1}}$~\citep{scholz2016ApJ}.
{The all-sky rate can be converted to the volumetric rate as $\mathcal{R}_{_{\rm FRB}}\sim 1.2\cdot10^4~{\rm Gpc^{-3}\,day^{-1}}$, assuming
that the Arecibo telescope can detect FRB~121102-like objects at distances out to $1$~Gpc.
Given the fact that $11$ bursts are detected in $0.6\,$day during the Arecibo survey, the total number of FRBs emitted by one repeating FRB source
throughout its lifetime
is roughly $N_{\rm tot}\sim 6.7\cdot10^4\,({\mathcal T}_{_{\rm FRB}}/10\,{\rm yr})$, where ${\mathcal T}_{_{\rm FRB}}$ is a typical lifetime of repeating FRB objects, which is
currently unknown. The birth rate of repeating FRB sources is estimated as $\mathcal{R}_{_{\rm FRB}}/N_{\rm tot}\sim 70 f_b^{-1}\,{\rm Gpc^{-3}\,yr^{-1}}\,({\mathcal T}_{_{\rm FRB}}/10\,{\rm yr})^{-1}$ (see \citealt{lu2016MNRAS,nicholl2017ApJ} for more detailed discussions taking the luminosity function into account).
Here $f_b\leq 1$ is a beaming factor of FRBs. }
If ${\mathcal T}_{_{\rm FRB}}\sim10~{\rm yr}$, this birth rate is compatible with the formation rate of  binary neutron stars 
(see, e.g., \citealt{kalogera2004ApJ,kim2015MNRAS} for the Galactic double neutron star systems and \citealt{wanderman2015MNRAS} for short GRBs). 

While the rate of rapidly rising optical transients is currently unknown, 
more systematic observational studies will allow us to reveal the event rate. This will enable us to test our scenario. 
Furthermore, if millisecond pulsars formed in neutron star binaries are the progenitor of the rapidly rising optical transients, 
bright radio pulses and persistent emission are expected to be associated with these events.  
Therefore the radio follow-up observations of these transients can confirm 
or rule out this scenario.
Also, successful detections will allow us to discover extragalactic pulsars with millisecond periods. 

Although the scenario has some testable predictions, we should note that the host property of FRB 121102 is so far against the possible connection between rapidly rising transients and repeating FRBs; the host galaxy of FRB 121101 is a dwarf-star-forming galaxy \citep{tendulkar2017ApJ}
while those of the rapidly rising transients are massive galaxies \citep{arcavi2016ApJ}. 
We definitely need more samples for this discussion too.

\section{Summary and discussion}\label{sec:con}
We studied optical counterparts of a new-born pulsar in double neutron star systems like PSR~J0737-3039A/B.
We considered the thermal emission arising from a pulsar wind embedded in the supernova ejecta. 
Given the ejecta mass, magnetic field's strength of the pulsar inferred from the PSR~J0737-3039A/B, and its initial spin, which
is inferred from the tidal synchronization of the progenitor star during the core He-burning phase, this emission is expected to have 
a peak bolometric luminosity of $\sim 10^{44}\,$erg/s and a rise time of $\sim 10$~days. 
In addition, the optical light curves have a long-lasting tail due to the photoelectric absorption of the ejecta to X-ray and UV photons
emitted by the pulsar wind nebula.  
These features are broadly consistent with those of the observed rapidly rising optical transients \citep{arcavi2016ApJ}.

There are several issues in our model.  
Regarding the pulsar model of ultra-stripped supernovae, one of the concerns is that the broad emission and absorption 
lines of H$\alpha$ are seen in the spectrum of PTF10iam, which are not expected for the explosion of ultra-stripped progenitors.
However, it could be explained by Si II, so  more detailed studies of the spectra
are needed \citep{arcavi2016ApJ}.
While we argued that tidal synchronization may lead to a fast-rotating pulsar as a remnant of the second supernova, initial magnetic 
fields are highly uncertain and stronger magnetic fields are possible. 
One should also note that the initial spin frequency, i.e., the initial spin down luminosity, of PSR~J0737-3039B
may be much lower than our estimate, depending on the mass loss history of the progenitor.
In order to address this issue, detailed studies based on the stellar evolution are needed. 

{We also discussed the possible connection between young binary neutron stars and 
FRBs. A small amount of the ejecta and high pulsar spin-down luminosity at the birth 
of the younger pulsars of binary neutron stars are in accord with the parameters for the repeating FRB 121102
 \citep{kashiyama2017ApJ}. Furthermore, the formation rate
of repeating FRBs seems consistent with that of binary neutron stars. These suggest that
young pulsars of binary neutron stars may produce FRBs.}

While this work focused on electromagnetic counterparts of new-born neutron star binaries, analogous arguments can be made for other nascent systems such as black hole-neutron star binaries and black hole binaries. In particular, a massive progenitor star that is tidally synchronized by the companion may lead to outflow-driven transients via the long-lasting accretion onto a new-born black hole at the second collapse~\citep{2017arXiv170207337K}.  
The spin evolution of compact binary progenitors is imprinted in the spin parameters of merging binary black holes, which can also be measured through gravitational-wave detections \cite{abbott2016PhRvX}.
Such measurements will shed lights on various questions on the formation scenario of compact binary objects that have been golden candidate sources of gravitational waves (\citealt{kushnir2016MNRAS,rodriguez2016ApJ,zaldarriaga2017,hotokezaka2017}). 


Apart from the emission of the ejecta powered by the pulsar wind, 
we expect there to be  significant non-thermal radiation from the pulsar wind nebula.
This non-thermal radiation has a broad spectrum from the radio to X and $\gamma$ rays.
These are also bright counterparts of a new-born pulsar in close double neutron star systems.
We will discuss this nebula emission and its detectability in a separate paper.
{It is also worthy to note that extragalactic binary pulsars can be a standard cosmological siren,
which may allow us to measure the expansion rate of the Universe \citep{seto2001PRL}. }


\acknowledgments
We thank Jason Hessels, Brian Metzger, Masaru Shibata, Anatoly Spitkovsky, and Ko Takahashi for useful discussion.
We are grateful to Iair Arcavi for providing us the observational data of rapidly rising transients. 
K. H. is supported by Flatiron Fellowship at the Simons Foundation.
K. K. acknowledges financial support from JST CREST.
The work of K. M. is supported by NSF grant No. PHY-1620777.

\end{document}